\documentclass{ws-p8-50x6-00}

\newcommand{\sla}{\!\!\!\!/ \,}

\begin{document}

\title{Gluon Condensate and Quark Propagation in the QGP}

\author{Andreas Sch\"afer}

\address{Institut f\"ur Theoretische Physik, Universit\"at Regensburg,
93040 Regensburg, Germany}

\author{Markus H. Thoma}

\address{ECT*, Villa Tambosi, Strada delle Tabarelle 286, 38050 Villazzano
(Trento), Italy}


\maketitle

\abstracts{A calculation of the thermal quark propagator 
is presented taking the gluon condensate above the critical 
temperature into account. The quark dispersion relation following
from this propagator is derived.} 

As an alternative method to lattice and perturbative QCD 
we suggest to include the gluon condensate into the parton propagators
\cite{ref4a}. 
In this way non-perturbative effects are taken into account within 
the Green functions technique. 

In the case of a pure gluon gas with energy density $\epsilon $ and 
pressure $p$ the gluon condensate can be related to the interaction 
measure $\Delta =(\epsilon -3p)/T^4$ via \cite{ref3}
\begin{equation}
\langle G^2 \rangle _T=\langle G^2  \rangle _0-\Delta T^4,
\label{e5}
\end{equation}
where $G^2\equiv (11\alpha _s/8\pi) : G^a_{\mu \nu}G_a^{\mu \nu}:$ 
and $\langle G^2  \rangle _0 = (2.5 \pm 1.0)\> T_c^4$ is the zero 
temperature condensate. Here $G_{\mu \nu}^a$ is the field strength tensor and 
$T_c$ the critical temperature of the phase transition to 
the quark-gluon plasma (QGP).

At zero temperature the
quark propagator containing the gluon condensate has been constructed
already \cite{ref5}. Here we will extend these calculations to finite
temperature.

\begin{figure}[t]
\centerline{\psfig{figure=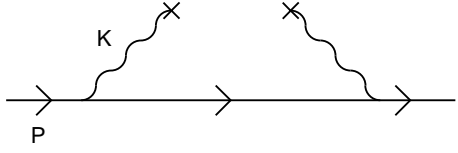,width=6cm}}
\caption{Quark self energy containing a gluon condensate.}
\end{figure} 

The full quark propagator in the QGP can be 
written by decomposing it according to its helicity eigenstates$\>$\cite{ref8}
\begin{equation}
S(P)=\frac{\gamma _0-\hat p\cdot \vec \gamma}
{2D_+(P)} + \frac{\gamma _0+\hat p\cdot \vec \gamma} {2D_-(P)},
\label{e12}
\end{equation}
where $D_\pm (P)=(-p_0\pm p)\> (1+a) - b$ and
\begin{eqnarray}
a & = & \frac{1}{4p^2}\> \left [tr\, (P\sla \Sigma ) - p_0\> tr\, 
(\gamma _0 \Sigma )\right ],\nonumber \\
b & = & \frac{1}{4p^2}\> \left [P^2\> tr\, (\gamma _0 \Sigma ) - p_0\> 
tr\, (P\sla \Sigma )\right ].
\label{e7} 
\end{eqnarray}

Using the imaginary time formalism and expanding the quark propagator
in Fig.1 for small loop momenta \cite{ref5}, 
i.e. $k\ll p$ and $k_0=2\pi inT=0$, we obtain
\begin{eqnarray}
a & = & -\frac{4}{3}\, g^2\, \frac{T}{P^6}\int \! \frac{d^3k}{(2\pi )^3}\,
\left [\left (\frac{1}{3}p^2-\frac{5}{3}p_0^2\right )k^2\, \tilde 
D_l(0,k)+ \left (\frac{2}{5}p^2-2p_0^2\right )k^2\, 
\tilde D_t(0,k)\right ], \nonumber \\
b & = & -\frac{4}{3}\, g^2\, \frac{T}{P^6}\int \! \frac{d^3k}{(2\pi )^3}\,
\left [\frac{8}{3}p_0^2\, k^2\, \tilde D_l(0,k)+ \frac{16}{15}p^2\, k^2\, 
\tilde D_t(0,k)\right ],
\label{e9} 
\end{eqnarray}
where $\tilde D_{l,t}$ are the longitudinal and transverse parts of the
non-perturbative gluon propagator at finite temperatur in Fig.1.   

The moments of the longitudinal and transverse gluon propagator in (\ref{e9}) 
are related to the chromoelectric and chromomagnetic condensates
via
\begin{eqnarray}
\langle {\bf E}^2\rangle _T & = & \langle :G_{0i}^aG_{0i}^a:\rangle _T
= 8T\> \int \frac{d^3k}{(2\pi )^3}\> k^2\> \tilde D_l(0,k),\nonumber \\
\langle {\bf B}^2\rangle _T & = & \frac{1}{2}\> 
\langle :G_{ij}^aG_{ij}^a:\rangle _T
= -16T\> \int \frac{d^3k}{(2\pi )^3}\> k^2\> \tilde D_t(0,k).
\label{e10} 
\end{eqnarray}
These condensates can be extracted from the expectation values of the
space- and timelike plaquettes $\Delta _{\sigma ,\tau}$ computed
on the lattice \cite{ref4}, using
\begin{eqnarray}
\frac{\alpha _s}{\pi }\> \langle {\bf E}^2 \rangle _T & = & \frac{4}{11}\>
\Delta _\tau\> T^4 - \frac{2}{11}\> \langle G^2\rangle _0,\nonumber \\ 
\frac{\alpha _s}{\pi }\> \langle {\bf B}^2 \rangle _T & = & -\frac{4}{11}\>
\Delta _\sigma\> T^4 + \frac{2}{11}\> \langle G^2\rangle _0.
\label{e11}
\end{eqnarray}

The quark dispersion relation \cite{ref8}, describing collective quark modes 
in the QGP in the presence of a gluon condensate, follows from $D_\pm (P)=0$. 
Using the lattice
results for the plaquette expectation values they have been 
determined numerically and are shown in Fig.2 for various temperatures. 

\begin{figure}[t]
\centerline{\psfig{figure=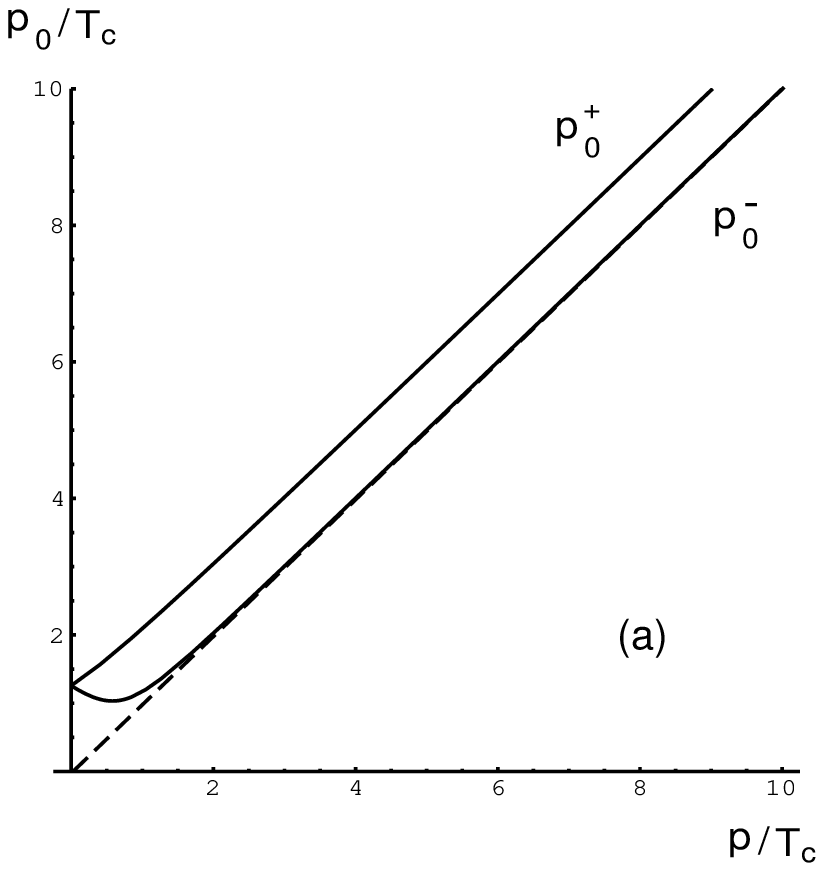,width=4cm}
\psfig{figure=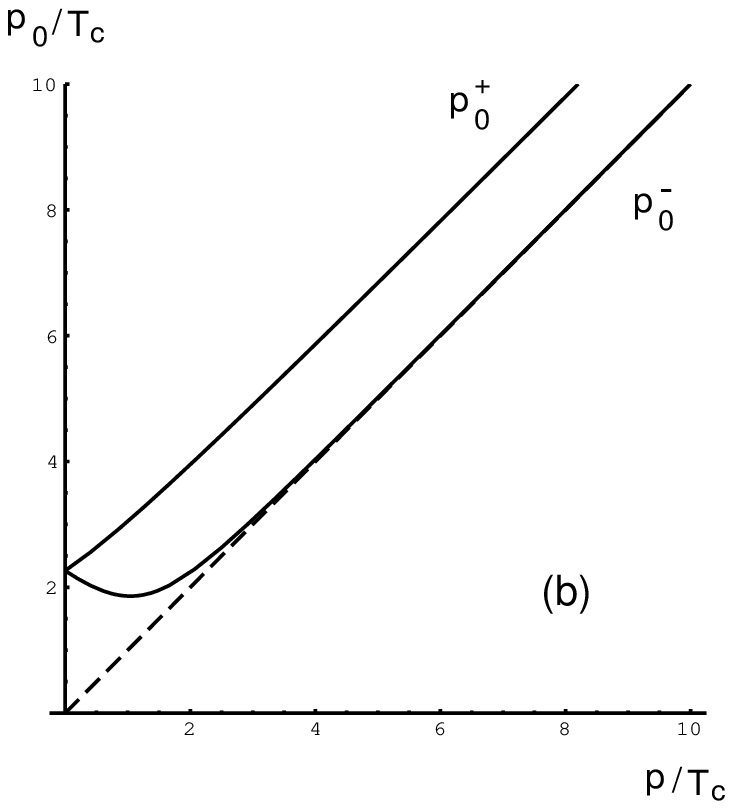,width=4cm}
\psfig{figure=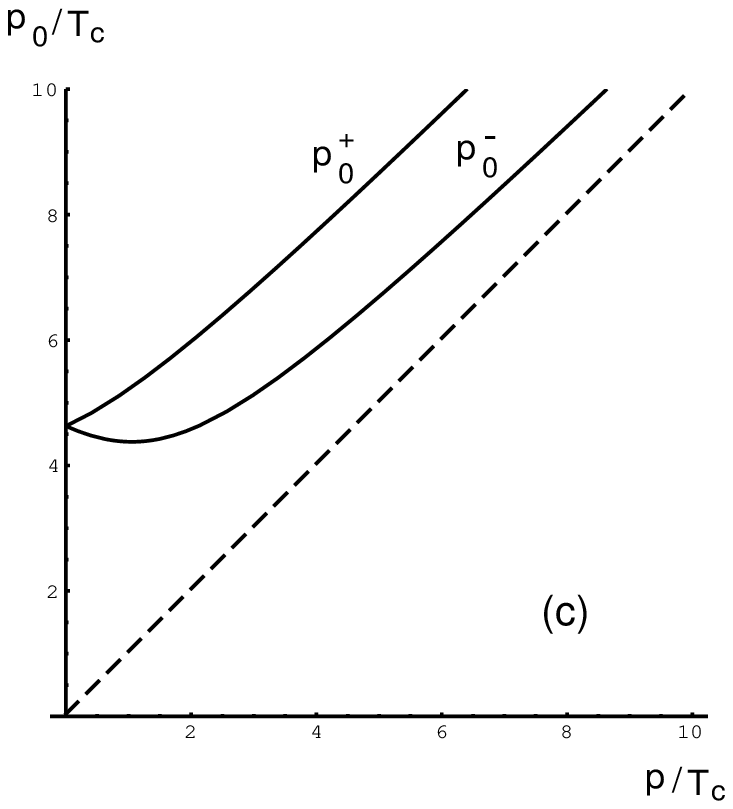,width=4cm}}
\caption{Quark dispersion relations at $T=1.1\, T_c$ (a), $T=2 \, T_c$ (b),
$T=4\, T_c$ (c) and dispersion relation of a non-interacting massless quark
(dashed lines).}
\end{figure}

The dispersions exhibit two massive quark modes. The upper branch comes 
from the solution of $D_+=0$ and the lower one from $D_-=0$. The lower
branch, showing a minimum, corresponds to a so-called plasmino, possessing 
a negative ratio of helicity to chirality, and is absent in the vacuum. 

At $p=0$ both modes start from a common effective quark mass, which is 
given by $m_{eff}=[(2\pi \alpha _s/3)\> 
(\langle {\bf E}^2\rangle _T + \langle {\bf B}^2\rangle _T )]^{1/4}$. 
In the temperature range $1.1 T_c <T< 4T_c$ we found approximately
$m_{eff}=1.15 \> T$. 

The qualitative picture of this quark dispersion relation is very similar
to the one found perturbatively in the HTL limit \cite{ref8}. The main 
difference is the different effective mass, which is given by 
$m_{eff}=gT/\sqrt{6}$ in the HTL approximation.

As a possible application of this effective quark propagator we mention the
computation of the photon and dilepton production rates from the QGP.
For this purpose the photon self energy using effective quark propagators
should be evaluated.

\end{document}